\documentclass[twocolumn,showpacs,amsmath,amssymb]{revtex4-1}

\usepackage{graphicx}
\usepackage{dcolumn}
\usepackage{bm}

\def\be{\begin{equation}}
\def\ee{\end{equation}}
\def\ba{\begin{array}}
\def\ea{\end{array}}
\def\bea{\begin{eqnarray}}
\def\eea{\end{eqnarray}}
\begin{document}

\title{Determination of electron screening potential of $^{6}$Li(p,$\alpha$)$^{3}$He reaction using Multi Layer Perceptron based neural network}
\author{ D. Chattopadhyay$^{1}$}
\email{dipayanchattopadhyay90@gmail.com}
\affiliation{$^1$Department of Physics, The ICFAI University Tripura, Agartala 799210, India}
\date{\today}

\begin{abstract}
\begin{description}
\item[Background] Understanding the nuclear reactions between light charged nuclei at sub-coulomb energy region holds significant importance in several astrophysical processes. Determination of the precise reaction cross-section within the astrophysically important Gamow range is difficult because of electron screening. Various polynomial fits, R-Matrix and Indirect Trojan horse method estimate much higher electron screening energies as compared to the adiabatic limit.
\item[Purpose] Obtain the bare astrophysical S-factor of $^{6}$Li(p,$\alpha$)$^{3}$He using Multi-Layer Perceptron based Artificial Neural Network based analysis and extract the electron screening energies. 
\item[Methods] Experimental S-factor of $^{6}$Li(p,$\alpha$)$^{3}$He, available in literature, are reanalyzed using the Multi-Layer Perceptron based Artificial Neural Network based algorithm to obtain the energy dependent astrophysical S-factor. Bare astrophysical S-factor is also calculated using the same Feed-forward Artificial Neural Network from the data range above 60 keV where the electron screening effect is expected to be negligible. Electron screening potential is then obtained by taking the ratio of total shielded S-factor with the bare S-factor.    
\item[Results and Conclusions] The electron screening potential obtained from the Present work through the Artificial Neural network based algorithm is found to be 220 eV. The extracted electron screening potential through the present analysis indicates that the Artificial Neural Network might be an alternative tools for estimation the electron screening potential involving light nuclei.      
\end{description}
\end{abstract}

\maketitle

\section*{Introduction}
Understanding the nuclear reactions mechanism between light charged nuclei at ultra low energies relevant to astrophysical energies is emerging as a burning topic in recent times. Measurement of precise cross-section is essential to understand the stellar evolution as well as abundances of various nuclides observed in the universes~\cite{rolfs88,adelberger11,engstler92,bertulani16,spitaleri16}.  Low energy nuclear reaction process involve transfer and capture processes, with the former primarily influenced by the nuclear force and the latter predominantly driven by the electromagnetic interaction. Determination of reaction cross-section at ultra low energies (much below the Coulomb barrier) of astrophysical interest requires considerable effort. Direct determination of cross-section in many cases is found to be very difficult or even impossible owing to small cross-section because of Coulomb repulsion of the interacting particles~\cite{rolfs88,broggini10}.  Extrapolation of direct cross-sections down to stellar energies requires extra theoretical effort. Different Indirect method has been evolved in recent times to overcome this problems~\cite{tribble14}. 

In this context, the low-energy cross sections governing reactions that produce or deplete lithium isotopes are pivotal. These cross sections are indispensable for addressing unresolved astrophysical questions, including the intricacies of Big Bang nucleosynthesis and the observed lithium depletion in celestial bodies like the Sun and other stars within our galaxy~\cite{spitaleri23}. The precise estimation of bare astrophysical cross-sections from experimental measurement is essential as they serve as key inputs in the stellar evolutionary codes. However, as the Gamow energy region is far below the Coulomb barrier, the cross-sections are very small and also very difficult to predict due to presence of electrons inside the target. Electrons inside the target nucleus effectively screen the coulomb potential at ultra low energies and thus reducing the height of the coulomb barrier. As a result the cross-section is enhanced by a factor exp$(\dfrac{U_{e}\pi\eta}{E})$~\cite{engstler92} and is given by 
\begin{equation}
\sigma(E)=\sigma_{b}(E)exp(\dfrac{U_{e}\pi\eta}{E})
\end{equation}

where, U$_{e}$ is the electron screening potential U$_{e}$(=${\dfrac{Z_{1}Z_{2}e^{2}}{R_{a}}}$ where, R$_{a}$ is atomic radius) arising from the interaction of the charged nuclei of the projectile with the electrons inside the target. In terms of astrophysical S-factor, the above expression can be written as  

\begin{equation}
S(E)=S_{b}(E)exp(\dfrac{U_{e}\pi\eta}{E})
\end{equation}

Similarly, in stellar plasma, ionized atoms are also surrounded by a sea of electrons within the Debye-Huckle radius and there also screening effect is important which depends on the temperature and density of the plasma. This plasma screening factor is different from the lab screening factor. Hence, it is essential to determine bare cross-section as accurate as possible by eliminating the electron screening effect from laboratory experiment. From the bare cross-section, the actual cross-section in stellar plasma can be determined through
\begin{equation}
\sigma_{pl}(E)=\sigma_{b}(E)f_{pl}
\end{equation}

where, f$_{pl}$(E) is the plasma screening enhancement factor and is related with the plasma screening factor by 
\begin{equation}
f_{pl}(E)=exp(\dfrac{U_{pl}\pi\eta}{E})
\end{equation}

U$_{pl}$(E) is the plasma screening potential and can be calculated within the framework of Debye-Huckel theory.  

Several attempts by different authors have been taken to understand the electron screening factor. Engstler {\it et al.}~\cite{engstler_92} have extracted screening potential for both the reactions $^{6}$Li(p,$\alpha$)$^{3}$He  using polynomial fit S(E)=a+b$E$+cE$^{2}$+dE$^{3}$. The sensitivity of the ratio of $E$ to U$_{e}$ has also been addressed by them. Extracted screening potential for both the reactions are found to be higher than the adiabatic limit (U$_{e}$=175 eV)~\cite{engstler92}. Later Cruz et al.~\cite{cruz08} have used lithium implanted targets and attempted to extract the screening potential for both $^{6}$Li(p,$\alpha$)$^{3}$He reaction. They have extracted potential as U$_{e}$=273$\pm$111 eV  for $^{6}$Li(p,$\alpha$)$^{3}$He. The values are lower than those obtained by Engstler {\it et al.}~\cite{engstler92} but still higher than the adiabatic limit. The uncertainty in the extracted electron screening potential is too large to make any feasible comment. In a systematic work by Barker~\cite{barker02} have also pointed out the higher value of lab screening potential with respect to the atomic adiabatic limit. 

Indirect Trojan Horse Method has been taken by Lamia et al.~\cite{lamia13} to obtain electron screening potential. They have also determined much higher potential than the predicted value of adiabatic limit. It is also interesting that the screening potential thus obtained for $^{6}$Li(p,$\alpha$)$^{3}$He (U$_{e}$=350$\pm$100 eV) is less than that for $^{7}$Li(p,$\alpha$)$\alpha$ (U$_{e}$=425$\pm$60 eV) which is also in contrary to the results obtained by Engstler et al.~\cite{engstler92} . Phenomenological R-Matrix method has been applied by Angulo et al.~\cite{angulo98} for $^{6}$Li(p,$\alpha$)$^{3}$He to obtain electron screening for both solid and gas target. Extracted electron screening potential is also found to higher for both cases.  

Artificial neural networks (ANN) serve as a powerful mathematical tool for estimating various values in science and technology, including nuclear physics studies. This method is particularly important in providing accurate results for highly nonlinear relationships between dependent and independent data variables. Recently, ANN has found application in numerous areas of nuclear physics, such as constructing consistent physical formulas for detector counts in neutron exit channel selection~\cite{akkoyun13}, determining one and two proton separation energies~\cite{athanassopoulos04}, developing nuclear mass systematic~\cite{athanassopoulos004}, determining ground state energies of nuclei~\cite{bayram14}, identifying impact parameters in heavy-ion collisions~\cite{david95,bass96,haddad97}, determining beta-decay energies~\cite{akkoyun14}, and estimating nuclear rms charge radius~\cite{akkoyun013}.

In this investigation, the Multi-Layer Perceptron based Feed-forward Artificial Neural Networks (ANN)~\cite{haykin99} using scikit-learn is utilized to estimate the bare astrophysical S-factor, total S-factor and subsequently the electron screening potential for $^{6}$Li(p,$\alpha$)$^{3}$He reaction. The experimental datas were extracted from existing literature~\cite{engstler92,engstler_92,gemeinhardt66,elwyn79,kwon89,shinozuka79,feidler67,cruz08,cruz05}. Through training the ANN with known experimental cross-section values for various combinations of beam, target, and bombarding energy, bare S-factor and total S-factor is successfully predicted. From the ratio of total S-factor with bare S-factor, the electron screening potential has been calculated. The details of Artificial Neural Network method will be discussed in next section followed by Calculations of Screening potentials and Conclusions.

\section*{Artificial Neural Network}
An artificial neural network (ANN) is a computational model that draws inspiration from the structure and functionality of biological neural networks found in the human brain. It serves as a crucial element in machine learning and artificial intelligence systems, aiming to emulate the information processing mechanism of the human brain. Artificial neural networks are constructed with neurons as their fundamental units. Each neuron receives inputs, undergoes computation, and generates an output. These neurons are arranged into layers, comprising an input layer, one or more hidden layers, and an output layer. In the case of a feed-forward neural network, information progresses unidirectionally from the input layer to the output layer without loops. The network includes multiple layers of nodes (neurons), featuring at least one hidden layer positioned between the input and output layers. Neurons establish connections through edges, each associated with a weight representing the strength of the connection. Additionally, a bias term is often employed to adjust the output of a neuron. Throughout the training phase, the weights and biases undergo adjustments to optimize the overall performance of the neural network. The activation function plays a crucial role in determining a neuron's output based on its input, introducing non-linearity and enabling the network to grasp intricate patterns. With a single output node, the network is well-suited for regression tasks, where the objective is to predict a continuous variable. During the feed-forward operation, input data progresses through the network layer by layer. Each neuron's output is computed by evaluating the weighted sum of its inputs and applying the activation function. Training a feed-forward neural network entails fine-tuning the weights and biases by considering the disparity between the predicted output and the actual target. This is typically accomplished through supervised learning using an optimization algorithm like gradient descent. The gradients of the loss concerning the weights and biases are computed, indicating how much the loss would change with adjustments to each weight and bias. The weights and biases are then modified in a direction that minimizes the loss, often employing optimization algorithms such as stochastic gradient descent or Adam. In the current analysis, a grid search will explore both `Adam' and `Stochastic Gradient Descent,' determining the most effective solver during the hyperparameter tuning process. These processes iterate for a specified number of training epochs (maxiter), progressively refining the model to enhance its predictive capabilities.

The subsequent discussion explores the mathematical framework of a straightforward feedforward neural network, specifically a Multi-Layer Perceptron (MLP) based for regression. I will start by breaking down the network layer by layer, and then proceed to introduce the training process, which incorporates gradient descent and adam optimization. 
\subsection*{Forward Pass:}
\begin{figure}[h]
\includegraphics[scale=0.32]{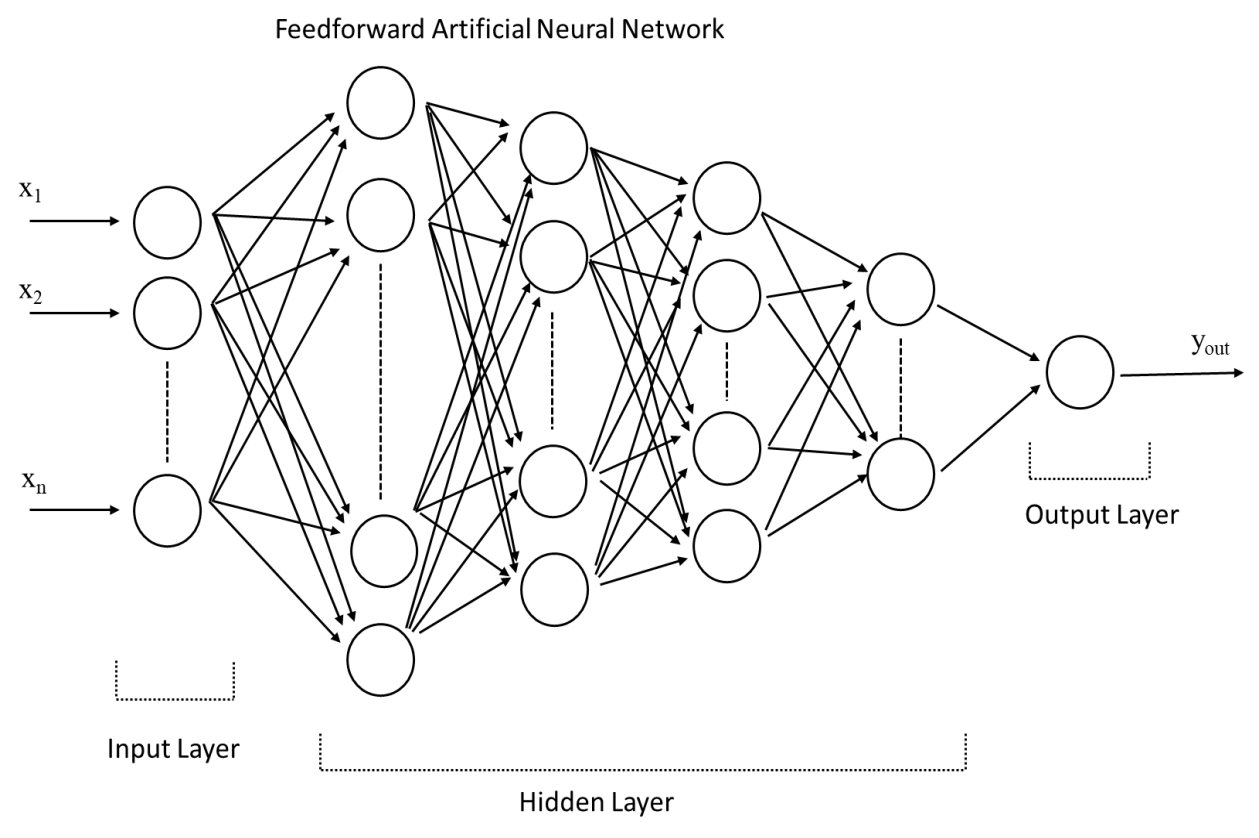}
\caption{\label{fig:ann} Typical architecture of Feed Forward Neural Network }
\end{figure}
\subsection{Input Layer:}
The neural network takes information about a single data point through a vector called X. If there are m features, X is like a vertical list with m rows and 1 column, often represented as:
\begin{align}
    X &= \begin{bmatrix}
           x_{1} \\
           x_{2} \\
           \vdots \\
           x_{m}
         \end{bmatrix}
  \end{align}   
Here, each x$_{i}$ represents a feature of the input data point. The information (x$_{1}$,x$_{2}$,...,x$_{m}$) passes to the nodes in the hidden layer. 

\subsection{Hidden Layer:}
For a single hidden layer consisting of $n$ neurons, the output Z$_{hidden}$ is calculated as:
\begin{equation}
Z_{hidden}=(W_{hidden}\cdot X+b_{hidden})
\end{equation} 
where,

\begin{figure*}
\centering\includegraphics[scale=0.50]{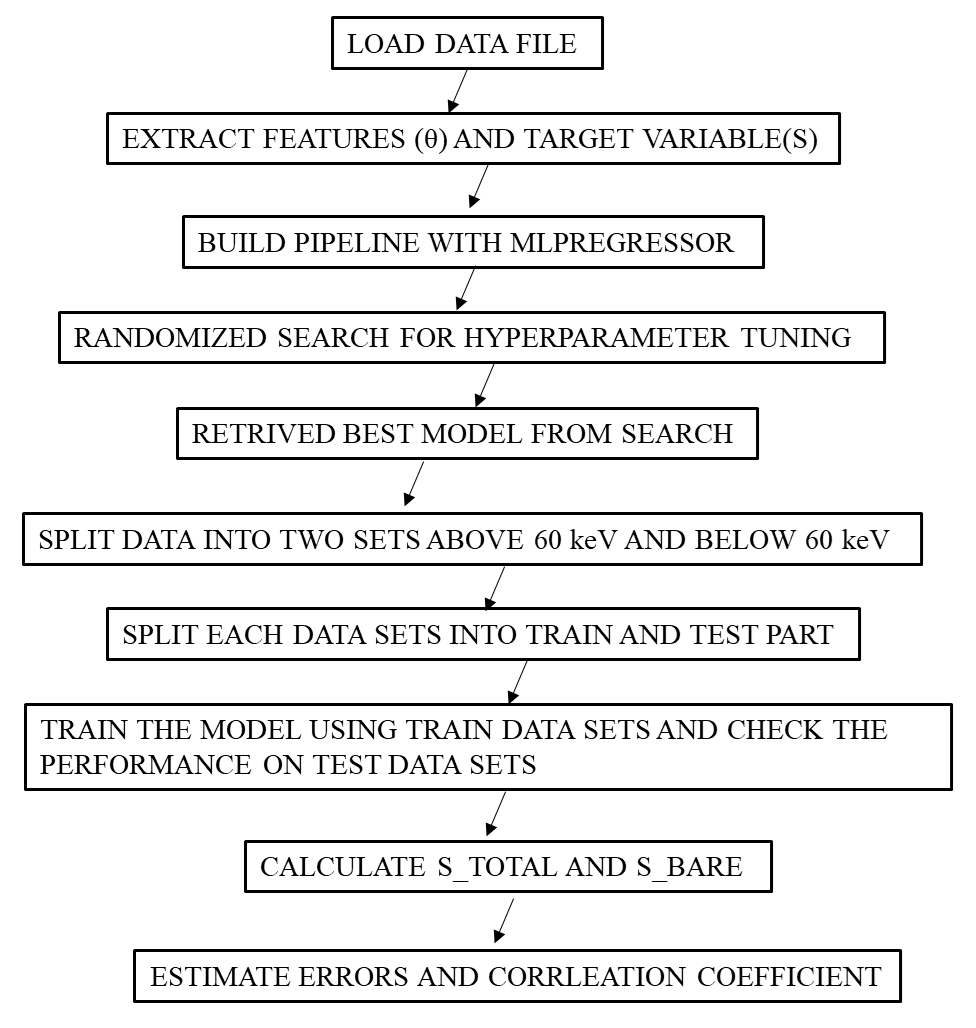}
\caption{\label{fig:flowchart} Flowchart of the ANN model}
\end{figure*}

W$_{hidden}$ is the weight matrix of size n $\times$ m.

b$_{hidden}$ is the bias vector of size n $\times$ 1.

After computing Z$_{hidden}$, it is passed through an activation function $\sigma$:
\begin{equation}
A=\sigma(Z_{hidden})
\end{equation}
Where, $A$ represents the output of the hidden layer. If the weighted sum of inputs to a neuron exceeds the threshold or activation layer, the neuron fires. In this analysis, the activation function for the hidden layers is being investigated with two choices: `ReLU' (Rectified Linear Unit) and `tanh' (Hyperbolic Tangent). The model will undergo training and evaluation using both activation functions to determine which one yields better performance for the provided data.

So, the overall computation can be written as:
\begin{equation}
A=\sigma(W_{hidden}\cdot X+b_{hidden})
\end{equation}

This represents the forward propagation through the first hidden layer of a neural network. The activation function $\sigma$ introduces non-linearity to the model, allowing it to learn complex patterns in the data.

In the present analysis four hidden layers are used in the neural network, the computation of the output at the first hidden layer Z$_{hidden1}$ with n$_{1}$ neurons in the layer and the subsequent hidden layers Z$_{hidden2}$ with n$_{2}$ neurons, Z$_{hidden3}$ with n$_{3}$ neurons and Z$_{hidden4}$ with n$_{4}$ neurons can be expressed as follows:

For the first hidden layer:
\begin{equation}
Z_{hidden1}=\sigma(W_{hidden1}\cdot X+b_{hidden1})
\end{equation} 
Here:

    W$_{hidden1}$ is the weight matrix of the first hidden layer with dimensions n$_{1}$ $\times$ m,
    
    b$_{hidden1}$ is the bias vector of the first hidden layer with dimensions n$_{1}$ $\times$ 1,

    $X$ is input vector with dimensions $m$ $\times$ 1, 
    
    $\sigma$ represents the activation function already mentioned above.

For the second hidden layer:
\begin{equation}
Z_{hidden2}=\sigma(W_{hidden2}\cdot Z_{hidden1}+b_{hidden2})
\end{equation} 

Here:

    W$_{hidden2}$ is the weight matrix of the second hidden layer with dimensions n$_{2}$ $\times$ n$_{1}$,
    
    b$_{hidden2}$ is the bias vector of the second hidden layer with dimensions n$_{2}$ $\times$ 1.

    Z$_{hidden1}$ is output of the first hidden layer with dimensions n$_{1}$ $\times$ 1.  

For the third hidden layer:
\begin{equation}
Z_{hidden3}=\sigma(W_{hidden3}\cdot Z_{hidden2}+b_{hidden3})
\end{equation} 

Here:

    W$_{hidden3}$ is the weight matrix of the second hidden layer with dimensions n$_{3}$ $\times$ n$_{2}$,
    
    b$_{hidden3}$ is the bias vector of the second hidden layer with dimensions n$_{3}$ $\times$ 1.

    Z$_{hidden2}$ is output of the first hidden layer with dimensions n$_{2}$ $\times$ 1.    

For the fourth hidden layer:
\begin{equation}
Z_{hidden4}=\sigma(W_{hidden4}\cdot Z_{hidden3}+b_{hidden4})
\end{equation} 

Here:

    W$_{hidden4}$ is the weight matrix of the second hidden layer with dimensions n$_{4}$ $\times$ n$_{3}$,
    
    b$_{hidden4}$ is the bias vector of the second hidden layer with dimensions n$_{4}$ $\times$ 1.

    Z$_{hidden3}$ is output of the first hidden layer with dimensions n$_{3}$ $\times$ 1.        

This process enhances the neural network's ability to learn hierarchical features and intricate representations from the input data.

\subsection{Output Layer:}
For regression, there is typically no activation function in the output layer, and the output Y is given by:
\begin{equation}
Y=W_{output} \cdot Z_{hidden4}+b _{output}
\end{equation} 

where, W$_{output}$ is a weight matrix of size 1 $\times$ n$_{4}$

Z$_{hidden4}$ is the output from the fourth hidden layer,

b$_{output}$ is the bias term for the output layer.

\subsection*{Training with Stochastic Gradient Descent:}

\subsection{Loss Function:}

The Mean Squared Error (MSE) is commonly used for regression:
\begin{equation}
MSE=\dfrac{\sum_{i=1}^{N} (\hat{Y_{i}}-Y_{i})^2}{N}
\end{equation} 
where,
N is the data points.
$\hat{Y_{i}}$ is the predicted output for data point $i$.
Y$_{i}$ is the actual output for data point i.
Normalized Mean Squared Error(NMSE) is defined as:
\begin{equation}
NMSE=\dfrac{MSE}{S_{max}-S_{min}}
\end{equation} 
where, S$_{max}$ and S$_{min}$ is the maximum and minimum value of astrophysical S-factor.
Furthermore, a regularization term (R) is incorporated into the loss function during training to counteract overfitting by imposing penalties on large weights in the model. In the case of L2 regularization, the regularization term is directly linked to the square of the weights:
\begin{equation}
R_{L2}=\dfrac{\alpha}{2}\sum_{i=1}^{n} W_{i}^{2}
\end{equation} 
Here, $\alpha$ represents the strength of the regularization term. Higher values of $\alpha$ result in stronger regularization.
\subsection{Back Propagation:}

Backpropagation (short for ``backward propagation of errors'') is a widely used supervised learning algorithm for training artificial neural networks. It is a gradient-based optimization algorithm that aims to minimize the error between the predicted output and the actual target output. The key steps in the backpropagation process for an artificial neural network (ANN) are as follows:

(1) Compute the gradient of the regularization term with respect to the output layer weights: $\dfrac{\partial R}{\partial W_{output}}$.

(2) Compute the gradient of the mean squared error with respect to the output layer weights: $\dfrac{\partial MSE}{\partial W_{output}}$.

(3) Compute the gradient of the mean squared error with respect to the output layer biases: $\dfrac{\partial MSE}{\partial b_{output}}$.

(4) Propagate the gradients through the network to compute the gradient with respect to the fourth hidden layer weights and biases: $\dfrac{\partial R}{\partial W_{hidden4}}$,$\dfrac{\partial MSE}{\partial W_{hidden4}}$, $\dfrac{\partial MSE}{\partial b_{hidden4}}$ .

(5) Propagate the gradients through the network to compute the gradient with respect to the third hidden layer weights and biases: $\dfrac{\partial R}{\partial W_{hidden3}}$,$\dfrac{\partial MSE}{\partial W_{hidden3}}$, $\dfrac{\partial MSE}{\partial b_{hidden3}}$ .

(6) Propagate the gradients through the network to compute the gradient with respect to the second hidden layer weights and biases: $\dfrac{\partial R}{\partial W_{hidden2}}$,$\dfrac{\partial MSE}{\partial W_{hidden2}}$, $\dfrac{\partial MSE}{\partial b_{hidden2}}$ .

(7) Propagate the gradients through the network to compute the gradient with respect to the first hidden layer weights and biases: $\dfrac{\partial R}{\partial W_{hidden1}}$,$\dfrac{\partial MSE}{\partial W_{hidden1}}$, $\dfrac{\partial MSE}{\partial b_{hidden1}}$ .

The process involves the backward propagation of gradients through each layer of the network. The chain rule is applied to calculate the gradients at each layer based on the gradients computed in subsequent layers. This iterative process helps update the weights and biases of the network, refining its parameters for improved performance over time.

\subsection{Update Weights and Biases:}

Update the weights and biases using the gradients and a learning rate $\eta$:
\begin{equation}
W_{output} \leftarrow W_{output}-\eta(\alpha\dfrac{\partial R}{\partial W_{output}}+\dfrac{\partial MSE}{\partial W_{output}})
\end{equation} 
\begin{equation}
b_{output} \leftarrow b_{output}-\eta \dfrac{\partial MSE}{\partial b_{output}}
\end{equation} 
\begin{equation}
W_{hidden4} \leftarrow W_{hidden4}-\eta(\alpha\dfrac{\partial R}{\partial W_{hidden4}}+\dfrac{\partial MSE}{\partial W_{hidden4}})
\end{equation} 
\begin{equation}
b_{hidden4} \leftarrow b_{hidden4}-\eta \dfrac{\partial MSE}{\partial b_{hidden4}}
\end{equation} 
\begin{equation}
W_{hidden3} \leftarrow W_{hidden3}-\eta(\alpha\dfrac{\partial R}{\partial W_{hidden3}}+\dfrac{\partial MSE}{\partial W_{hidden3}})
\end{equation} 
\begin{equation}
b_{hidden3} \leftarrow b_{hidden3}-\eta \dfrac{\partial MSE}{\partial b_{hidden3}}
\end{equation} 
\begin{equation}
W_{hidden2} \leftarrow W_{hidden2}-\eta(\alpha\dfrac{\partial R}{\partial W_{hidden2}}+\dfrac{\partial MSE}{\partial W_{hidden2}})
\end{equation} 
\begin{equation}
b_{hidden2} \leftarrow b_{hidden2}-\eta \dfrac{\partial MSE}{\partial b_{hidden2}}
\end{equation} 
\begin{equation}
W_{hidden1} \leftarrow W_{hidden1}-\eta(\alpha\dfrac{\partial R}{\partial W_{hidden1}}+\dfrac{\partial MSE}{\partial W_{hidden1}})
\end{equation} 
\begin{equation}
b_{hidden1} \leftarrow b_{hidden1}-\eta \dfrac{\partial MSE}{\partial b_{hidden1}}
\end{equation} 
where, $R$ represents the regularization term added to the loss function during training. 
\begin{table} [h]
\caption{Summary of Experimental Data used in the $\Delta$E$_{cm}$ range from 10.7 keV to 927.4 keV with sources.}
\label{tab1}
\vspace*{0.05cm}
\begin{tabular}{cccc}\hline \hline
&&&\\
Data Used & Data Sources & References \\
&  &  &  \\
&&& \\ \hline
&&& \\
Engstler {\it et. al}&{\sc EXFOR}&~\cite{engstler92,engstler_92}\\
&&& \\
Gemeinhardt {\it et. al}&{\sc EXFOR}&~\cite{gemeinhardt66}\\
&&& \\
Elwyn {\it et. al}&{\sc EXFOR}&~\cite{elwyn79}\\
&&& \\
Kwon {\it et. al}&{\sc EXFOR}&~\cite{kwon89}\\
&&& \\
Shinozuku {\it et. al}&{\sc EXFOR}&~\cite{shinozuka79}\\
&&& \\
Feidler {\it et. al}&{\sc EXFOR}&~\cite{feidler67}\\
&&& \\
Cruz {\it et. al}&{\sc EXFOR}&~\cite{cruz08}\\
&&& \\
Cruz {\it et. al}&{\sc EXFOR}&~\cite{cruz05}\\
&&& \\ \hline
\end{tabular}
\end{table}

\begin{table} [h]
\caption{Bare astrophysical S-factor at zero energy(S$_{b}$(0)) for the $^{6}$Li(p,$\alpha$)$^{3}$He reaction from different sources}
\label{tab2}
\vspace*{0.05cm}
\begin{tabular}{ccccc}\hline \hline
&&&&\\
S$_{b}$(0) & $\Delta$S$_{b}$(0) & Method & References \\
(MeV-b)&(MeV-b)&   &  \\
&&&& \\ \hline
&&&& \\
3.09&$\pm$1.23&Polynomial extrapolation&~\cite{engstler92}\\
&&&& \\
3.56&$---$&Polynomial extrapolation&~\cite{barker02}\\
&&&& \\
3.00&$\pm$0.19&THM&~\cite{tumino03}\\
&&&& \\
3.52&$\pm$0.08&Polynomial extrapolation&~\cite{cruz08}\\
&&&& \\
3.63&$\pm$0.13&Polynomial extrapolation&~\cite{wang11}\\
&&&& \\
3&$\pm$0.50&Compilation&~\cite{xu13}\\
&&&& \\
3.44&$\pm$0.35&THM&~\cite{lamia13}\\
&&&& \\
3.37&$\pm$0.50&R-Matrix&~\cite{spitaleri23}\\
&&&& \\
3.36&$\pm$0.12&ANN&Present Work\\
&&&& \\ \hline
\end{tabular}
\end{table}
This process is repeated for maximum iterations until convergence.
The training with Stochastic Optimization using Adam(Adaptive Moment Estimation) has been described in ~\cite{kingma14}. 
\begin{table} [h]
\caption{The Electron Screening potential U$_{e}$(Lab) for the $^{6}$Li(p,$\alpha$)$^{3}$He reaction from different sources.}
\label{tab3}
\vspace*{0.1cm}
\begin{tabular}{ccccc}\hline \hline
&&&&\\
Sl. No. & U$_{e}$(Lab) & $\Delta$U$_{e}$(Lab) & Method & References\\
& (MeV-b) & (MeV-b) &  &\\
&&&& \\ \hline
&&&& \\
1&470&$\pm$150& &~\cite{engstler92}\\
&&&& \\
2&440&$\pm$150& &~\cite{engstler92}\\
&&&& \\
3&260&$---$&Polyn. Fit&~\cite{barker02}\\
&&&& \\
4&450&$\pm$100&THM&~\cite{tumino03}\\
&&&& \\
5&237&$\pm$111& &~\cite{cruz08}\\
&&&& \\
6&310&$\pm$109& &~\cite{wang11}\\
&&&& \\
7&218&$\pm$38& &~\cite{wang11}\\
&&&& \\
9&355&$\pm$100&THM &~\cite{lamia13}\\
&&&& \\
10&290&$\pm$75&R-MATRIX &~\cite{spitaleri23}\\
&&&& \\
11&220&$---$& ANN &Present Work\\
&&&& \\ \hline
\end{tabular}
\end{table}
This is a simplified overview, and the actual scikit-learn implementation may involve additional details. 
\begin{figure}[h]
\includegraphics[scale=0.45]{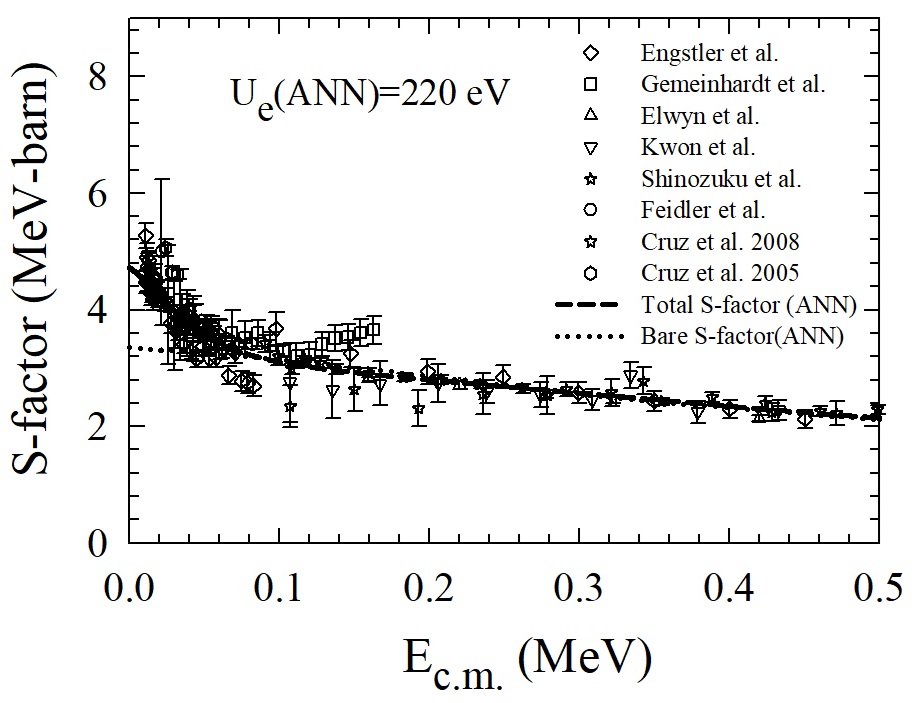}
\caption{\label{fig:screening} Comparison of experimental data with the shielded-nucleus total S-factor and bare S-factor obtained from present calculation based of ANN model.}
\end{figure}
\begin{figure}[h]
\includegraphics[scale=0.45]{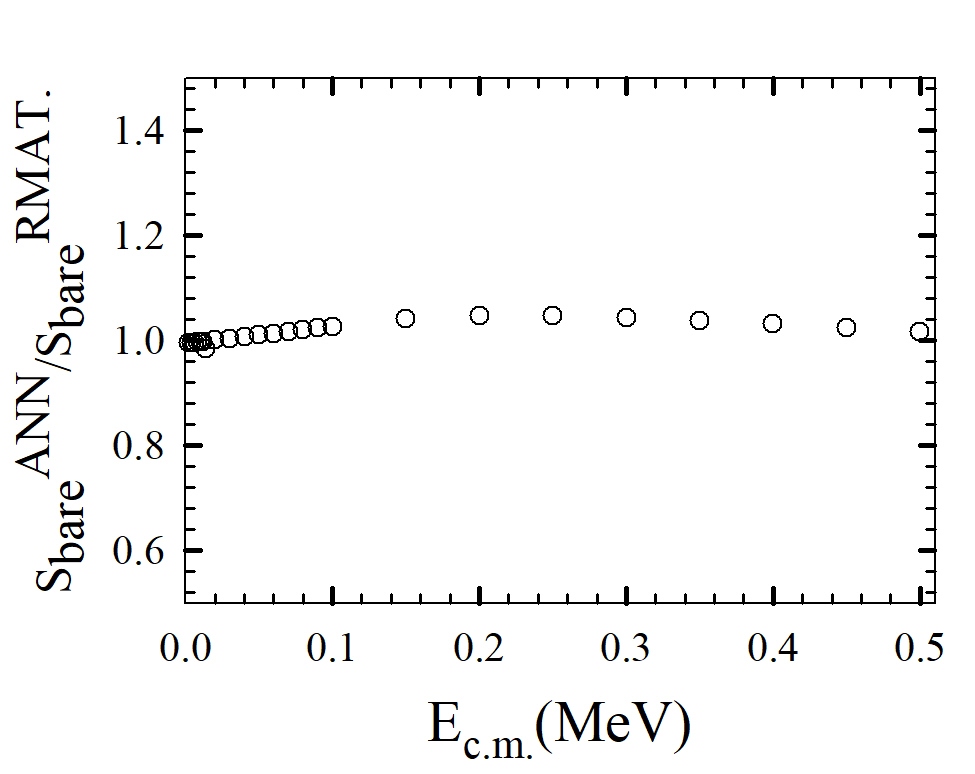}
\caption{\label{fig:sbare_rmat} Comparison of Bare S factor obtained from the present ANN model with that obtained from R-Matrix calculations.}
\end{figure}
\begin{figure}[h]
\includegraphics[scale=0.45]{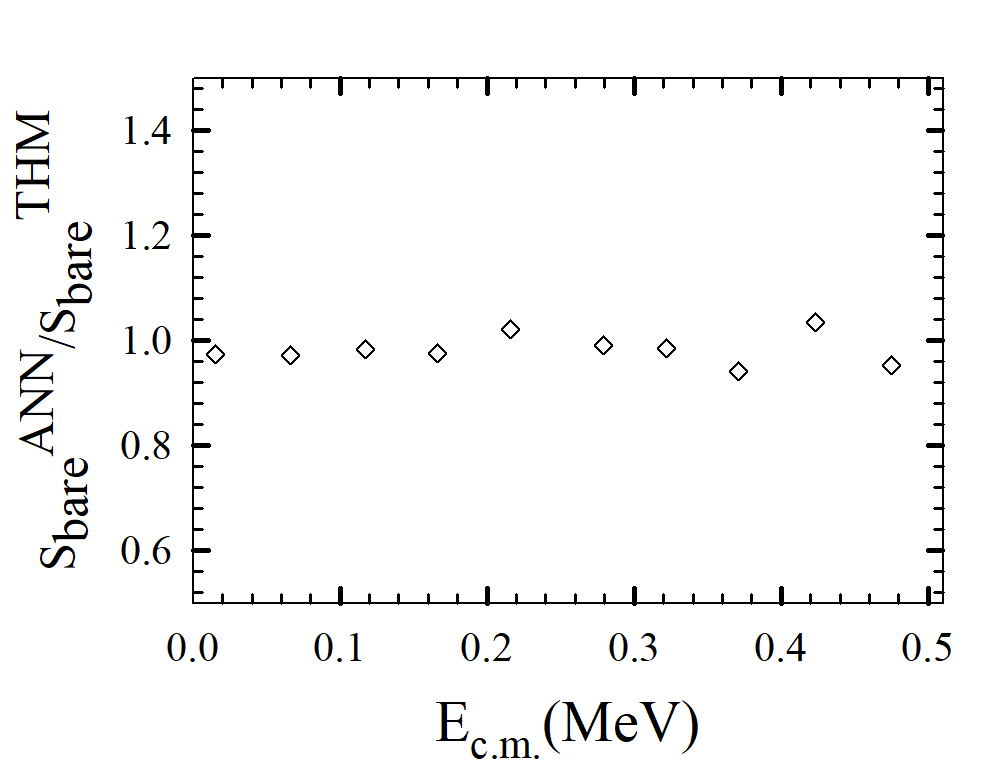}
\caption{\label{fig:sbare_thm} Comparison of Bare S factor obtained from the present ANN model with that obtained from the THM calculations.}
\end{figure}
\begin{figure}[h]
\includegraphics[scale=0.45]{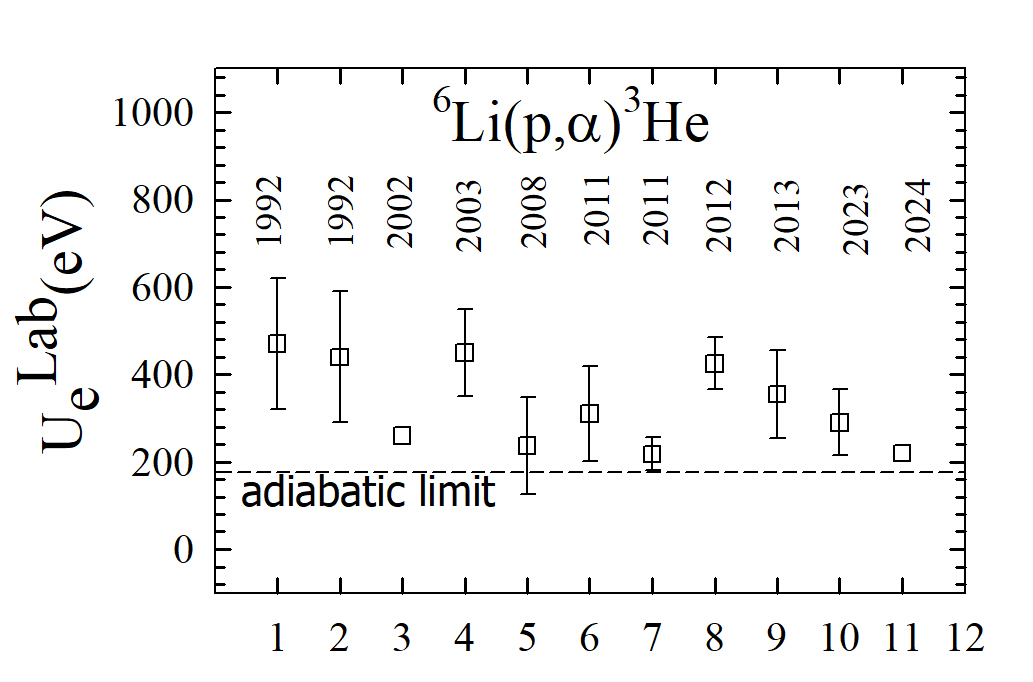}
\caption{\label{fig:Ue_year} Comparison of Electron screening potential obtained from the present calculation with the already available published data.}
\end{figure}
\section*{Calculations of Electron Screening Potential}
Multi Layer Perceptron based ANN algorithm has been used to find the S-factor from the available experimentally measured literature data. Those data sets are chosen in present calculations which are experimentally well established and have better accuracy. Reliable experimental data are crucial as it will contribution to the robustness of the present ANN model calculations, that can help in drawing meaningful conclusions. Table~\ref{tab1} summarizes the different data sets used for the present analysis. The data sets are directly taken from {\sc EXFOR} database. Some datasets in the existing literature exhibit notable discrepancies with the majority of previous findings are excluded in the present analysis. The utilized data sets cover an energy range in which both electronic screening effects are expected to be present(below $\approx$ 60 keV) and in which they are not expected to be present(above $\approx$ 60 keV). The collecting dataset has then been splitted randomly into train and test part. The train part contains 70$\%$ and test part contains 30$\%$ of total data set. Model is trained using the train dataset. During the initial phase of the study, following numerous trials, the optimal number of hidden layer neurons was identified and employed in a consolidated four-in-one layer which demonstrated the most effective outcomes for the present problem. Present analysis employs RandomizedSearchCV with a comprehensive grid search for optimal hyperparameters. Hyperparameters play a crucial role in machine learning algorithms by controlling the learning process and shaping the final model's behavior. Optimized hyperparameters are used to obtain the best correlation coefficient. Higher correlation coefficient indicates the lower uncertainty and vice versa.  In the present calculations, one input layer, four hidden layers with size (128,64,32,16) and one output layer is used after obtaining this configuration from several trials for optimization. The number of hidden layer and hidden units are so chosen to get the minimum loss as much as possible. The maximum iteration is used as 5000. In the Adam algorithm, the initial learning rate is 0.001 and decay constants are 0.9 and 0.999. Once the model are trained, the same set of Hyperparameters are used to predict the performance of the model on test data-set.  In the second part, data's are above 60 keV are used to find the bare S-factor using the same ANN code with same hyperparameters. Once the bare S-factor is obtained the electron screening potential is then obtained by taking the ratio of the total S-factor with the bare S-factor. The total S-factor and bare S-factor obtained from the present ANN based model is shown in Fig.~\ref{fig:screening}. Bare astrophysical S-factor at zero energy calculated from the present model is 3.36 Mev-barn and it is tabulated in Table~\ref{tab2} and compared with the value obtained by the different methods in Literature. The values obtained from the present calculation is in excellent agreement with that obtained from the R-MATRIX calculation~\cite{spitaleri23} and THM calculation~\cite{lamia13} by taking into account the systematic uncertainty. The variation of bare astrophysical S-factor with that obtained from R-MATRIX and THM has been compared in Fig.~\ref{fig:sbare_rmat} and Fig.~\ref{fig:sbare_thm} respectively. It has been observed from the comparison that the present calculations reasonably matches with the well established R-MATRIX and THM calculation. Electron screening potential obtained from the present calculation (U$_{e}$=220 eV) has been tabulated in Table~\ref{tab3} and compared with the electron screening potential obtained from the other methods as shown in Fig.~\ref{fig:Ue_year}. The electron screening potential thus obtained from the present ANN based model nicely matches with the other methods within the quoted uncertainty. This results indicates that Multi-Layer Perceptron based Artificial Neural Network model can be used an alternative tools to find the the electron screening potential in light charged particle induced reactions of astrophysical interest.   

\section*{Conclusions}
In this study, the electron screening potential is obtained for $^{6}$Li(p,$\alpha$)$^{3}$He reactions by using the Multi-Layer Perceptron based Artificial Neural Network based algorithm for the first time. From the present analysis, the electron screening potential is obatined as 220 eV. The electron screening potential matches resonably well with other welll estrablished techniques like R-MATRIX calculation and THM method indicating the ANN method as an alternative powerful tool for extraction of such astrophysical quantity. The variation of the bare S-factor with R-MATRIX and THM calculation also supports the claim.

\end{document}